\documentclass[twocolumn]{revtex4}

\usepackage{amssymb}
\usepackage{graphicx}
\usepackage{amsmath}
\usepackage{amsfonts}
\usepackage{natbib}
\usepackage{slashbox}

\begin{document}

\title{Entanglement Swapping Model of DNA Replication}
\author{Onur Pusuluk$^{1}$}
\author{Cemsinan Deliduman$^{2}$}
\affiliation{$^1$ Engineering Physics Department, Faculty of Sciences and Letters\\
\.{I}stanbul Technical University, Maslak 34469, \.{I}stanbul, Turkey}
\affiliation{$^2$ Department of Physics, Faculty of Sciences and Letters\\
Mimar Sinan Fine Arts University, Be\c{s}ikta\c{s} 34349, \.{I}stanbul, Turkey}

\begin{abstract}
Molecular biology explains function of molecules by their
geometrical and electronical structures that are mainly determined
by utilization of quantum effects in chemistry. However, further
quantum effects are not thought to play any significant role in the
essential processes of life. On the contrary, consideration of
quantum circuits/protocols and organic molecules as software and
hardware of living systems that are co-optimized during evolution,
may be useful to overcome the difficulties raised by biochemical
complexity and to understand the physics of life. In this sense, we
review quantum information-theoretic approaches to the process of
DNA replication and propose a new model in which 1) molecular
recognition of a nucleobase is assumed to trigger an intrabase
entanglement corresponding to a superposition of different tautomer
forms and 2) pairing of complementary nucleobases is described by
swapping intrabase entanglements with interbase entanglements. We
examine possible biochemical realizations of quantum
circuits/protocols to be used to obtain intrabase and interbase
entanglements. We deal with the problem of cellular decoherence by
using the theory of decoherence-free subspaces and subsystems.
Lastly, we discuss feasibility of the computational or experimental
verification of the model and future research directions.
\end{abstract}

\maketitle

\section{Introduction}

According to the central dogma of molecular biology, genetic
information stored in double-stranded DNA (dsDNA) is duplicated by
replication of two strands independently. At each step of the
replication, enzyme DNA polymerase (DNA\emph{pol}) first recognizes
the nucleobase (N = \{A, T, G, C\}) of the template DNA strand.
Then, it finds complementary of this base (\={N} = \{\={A}=T,
\={T}=A, \={G}=C, \={C}=G\}) from the surrounding environment and
facilitates pairing of these bases through two or three interbase
hydrogen bonds. A new dsDNA is synthesized from an existing
single-stranded DNA (ssDNA) by successive pairings of the all
nucleobases in this way.

Conformational changes occurring within DNA\emph{pol} at each stage
of the replication were demonstrated in detail by crystallization
experiments \citep{KE-1}. Also, all possible interactions between
amino acid side chains (that are likely to be found in the active
site of DNA\emph{pol}) and unpaired nucleobases were obtained by
quantum chemical calculations \citep{KE-9}. However, there are still
some unclear points about the relation of the high fidelity of
replication process with the base recognition, searching, and
pairing mechanisms \citep{KE-1}. Since active site of DNA\emph{pol}
that contribute to these mechanisms has a particularly complex
structure involving a lot of amino acids \citep{KE-1}, both
experiments and quantum chemical calculations are still insufficient
to clarify these mysteries. Thus, until the development of more
sensitive setups and more sophisticated calculations, information
processing models could be useful tools for a better understanding
of DNA replication.

During the DNA replication, newly synthesized strands elongate with
a rate 3,000 nucleotide per minute in humans \citep{K-1} and 30,000
nucleotide per minute in bacteria \citep{K-2}. Neither DNA binding
nor nucleotide binding to the DNA\emph{pol} limits this rate, they
are very fast steps \citep{KE-1, KE-23, KE-24}. Also, replication
without proofreading and repair mechanisms occurs with an error rate
of $10^{-4}$ to $10^{-6}$ per nucleotide \citep{K-2}. Such an
accuracy can be within the constraints of quantum coherent
information processing. Estimations based on both theoretical models
\citep{decoh_1} and experimental data \citep{decoh_4, decoh_5} give
sufficiently long decoherence times for the coding nucleobase
protons of dsDNA \citep{KE-17}. Thus, quantum information processing
descriptions are expected to be explanatory models of DNA
replication.

To understand the underlying mechanisms of DNA replication several
quantum information processing models were proposed. For example,
\citet{KE-2} formulated nucleotide selection from surrounding
environment as an unsorted database search. He examined the
pertinence of Grover's algorithm \citep{KE-7} to give an explanation
for the number of deoxyribonucleotide types used in dsDNA. Although
he successfully modeled base pairing as oracle of the algorithm
\citep{KE-2} and associated this model with the evolution of the
triplet genetic code \citep{KE-14}, initiation of the search in his
model requires the symmetric quantum superposition of four disparate
nucleotides which is not quite possible. Wave analogue \citep{KE-3}
of this quantum search algorithm in which symmetric superposition
state is replaced by the center-of-mass mode is more realistic for
enzyme activity. However, if this version of the algorithm
\citep{KE-3} is adopted for the activity of DNA\emph{pol}, each base
pairing should begin with the loading of DNA\emph{pol} with four
different free nucleotides before attempting to bind to DNA which is
contrary to the present knowledge \citep{KE-1, KE-23, KE-24}.
\begin{figure}[h]
\centering
\includegraphics[width=0.33\textwidth]{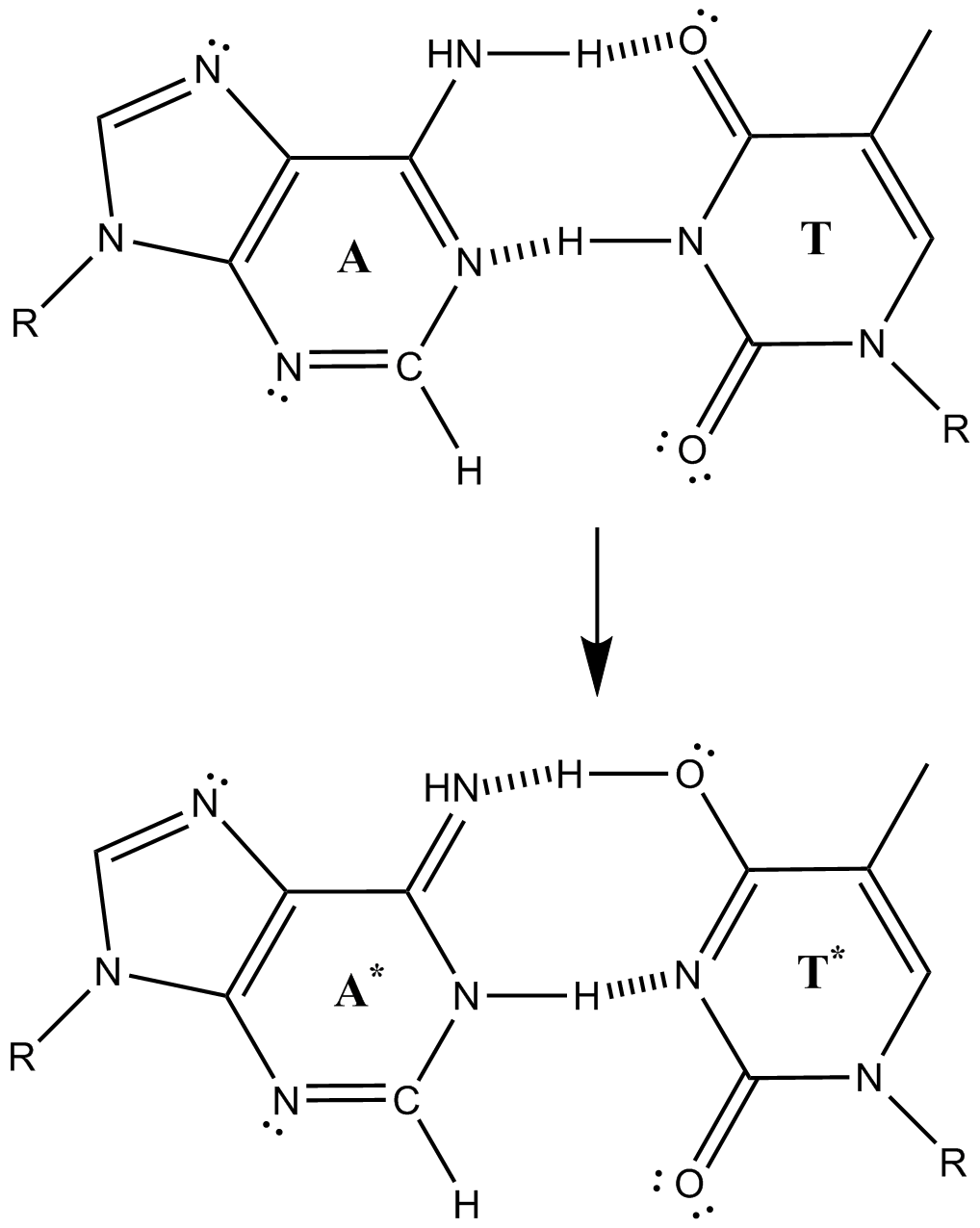}
\caption{A$\cdot$T$\rightarrow$A$^{*}$$\cdot$T$^{*}$ tautomeric
transition which can be observed \citep{HBPT-4, HBPT-3, HBPT-6,
HBPT-9, HBPT-11} in dsDNA.} \label{F-AT}
\end{figure}
\begin{figure}[h]
\centering
\includegraphics[width=0.33\textwidth]{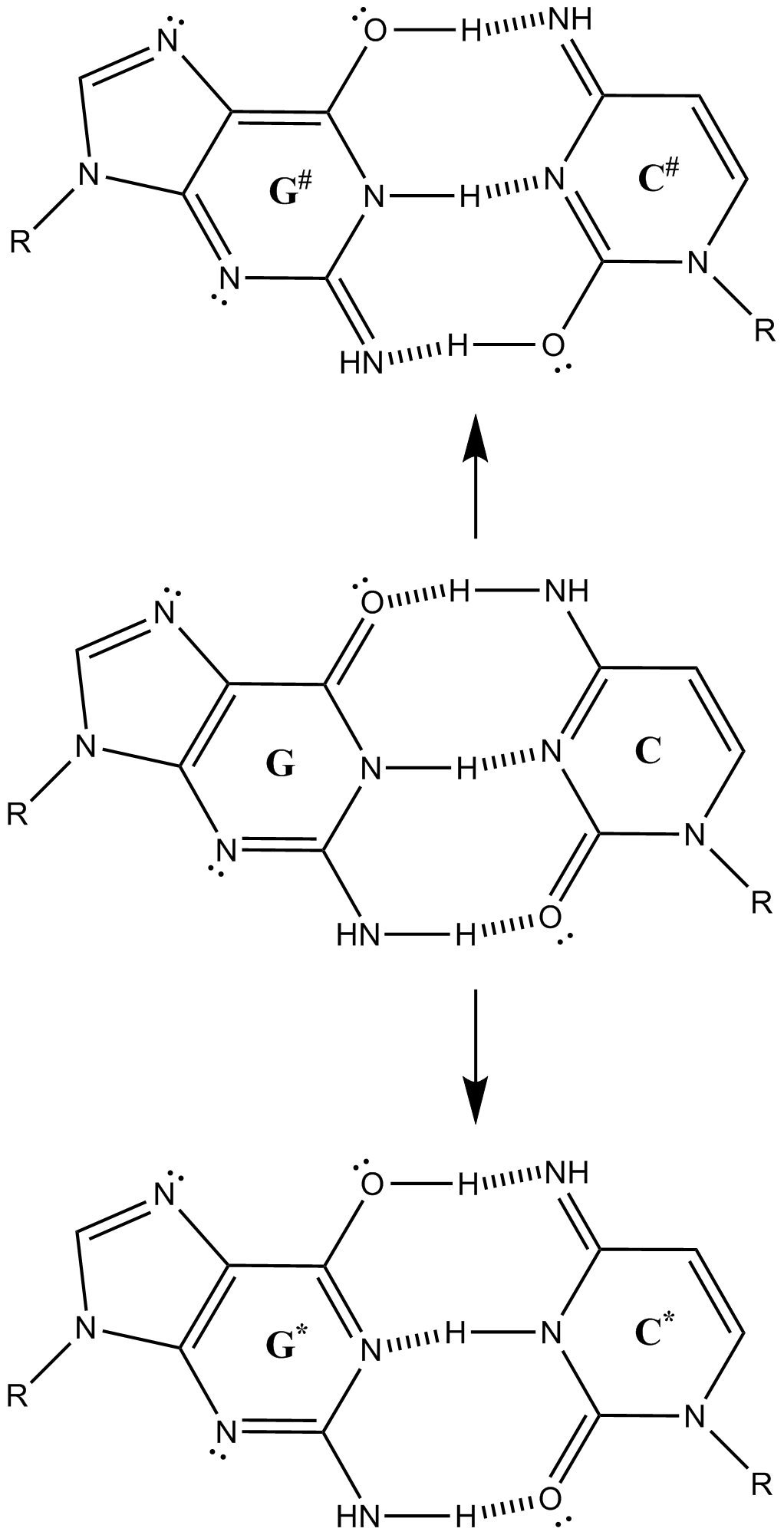}
\caption{G$\cdot$C$\rightarrow$G$^{\sharp}$$\cdot$C$^{\sharp}$ and
G$\cdot$C$\rightarrow$G$^{*}$$\cdot$C$^{*}$ tautomeric transitions
which can be observed \citep{HBPT-2, HBPT-4, HBPT-8, HBPT-10} in
dsDNA.} \label{F-GC}
\end{figure}

\begin{figure}[h]
\centering
\includegraphics[width=0.35\textwidth]{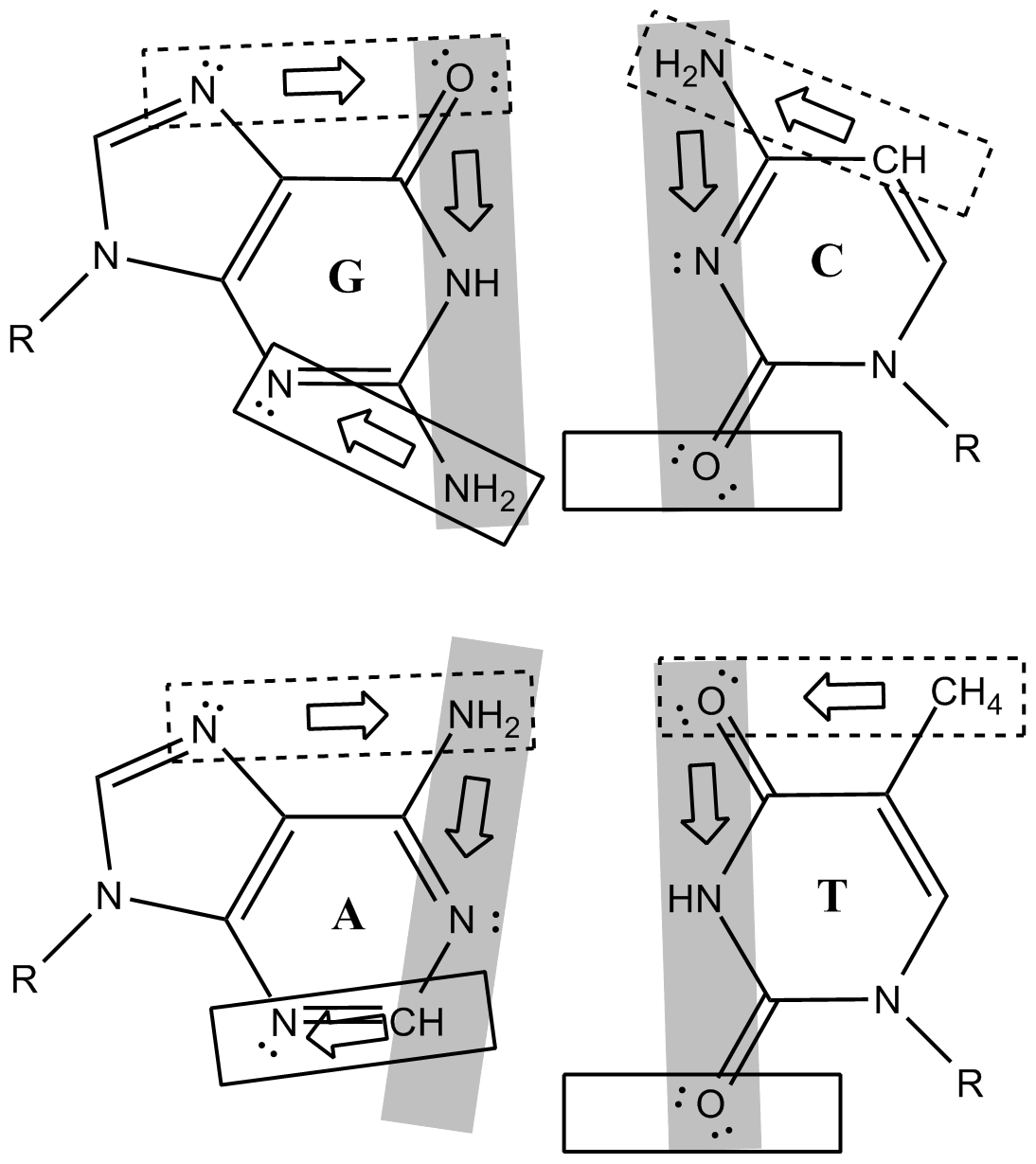}
\caption{Parts of the nucleobases: atoms can be grouped according to
region they will be found in DNA. Hoogsteen ($\leftrightarrow$major
groove), Watson-Crick ($\leftrightarrow$pairing plane), and Sugar
($\leftrightarrow$minor grove) edges are indicated respectively by
dashed, filled, and plain boxes. Arrows inside the boxes show the
order of the qubits used in qubit representation.} \label{F-parts}
\end{figure}

Recently, \citet{KE-4, KE-5} modeled base recognition mechanism in
replication (and transcription) to understand time-dependent DNA
mutations and A$\cdot$T richness of DNA. To explain the stability of
base pairs, he assumed that interbase hydrogen bonds are rearranged
by sequential intermolecular and intramolecular proton tunnelings.
In this assumption, interbase tunnelings turn bases into their
unusual tautomers ($\text{N}^*$ and $\text{N}^\sharp$ in Figures
\ref{F-AT}, \ref{F-GC}) pair-by-pair. Then, intramolecular
tunnelings introduce coherent superposition states in which enol and
imine protons of unusual tautomers are shared between two electron
lone pairs that belong to a single atom. In the recognition step of
the model, enzyme transcriptase (RNA-dependent DNA\emph{pol}) makes
quantum measurements on the coherent states of protons that are
present on Watson-Crick ($WC$) edge (Figure \ref{F-parts}). Although
this model is compatible with molecular genetic transcription data
of bacteriophage T4, some possible results of the transcriptase
measurements, such as the decohered states corresponding to
tautomers G$^{\sharp}_{002}$ and G$^{\sharp}_{000}$ \citep{KE-4,
KE-5} do not generate information for any usual tautomer form:
technically, qubit representation of a nucleobase was considered as
the tensor product of states corresponding to the presence or
absence of each $WC$ edge proton. Then, Hilbert space should be
8-dimensional, but four bases states, that do not correspond to
common tautomers, give no meaningful information to the enzyme.
Therefore, in such situations where the result of the measurement is
one of these spurious states, enzyme can not recognize the
nucleotide base of DNA. This expresses an efficiency problem in both
recognition and searching mechanisms.

\section{Entanglement Swapping Model of Replication}

Estimation of long decoherence times for the $WC$ edge protons
\citep{KE-17} suggests that nontrivial quantum effects can be
involved in DNA replication. It is easy to show that the fastest
quantum search algorithm \citep{KE-7, KE-25} is only four times
faster than the slowest classical search algorithm for the case of
searching complementary nucleotides of template bases. Therefore, if
quantum coherence is maintained during replication, this should have
been evolved to increase not only the speed, but also the accuracy.
In this sense, our model is motivated to investigate quantum effects
increasing both speed and accuracy together in the DNA replication.

\subsection{Intrabase and Interbase Entanglements}

During the DNA replication process free nucleotide binds to a
solvent exposed pocket within DNA\emph{pol} before base pairing,
whereas template base is flipped out of the helix axis and into the
active site \citep{KE-1}. Additionally, it is theoretically known
that interaction with water molecules can induce transitions to rare
tautomer forms \citep{KE-26}. Such higher energetic states can also
be mediated by interactions with carboxylate and sodium ions
\citep{KE-27} which are likely to be found both in the solvent
exposed pocket and active site of the enzyme. Thus, tautomeric
transitions are likely in both incorporated nucleobase and template
base after recognition and before base pairing.

In this work, molecular recognition of a nucleobase is assumed to
trigger a superposition of different tautomer forms, i.e.
$|\text{N}\rangle_{WC,I} \rightarrow |\text{N}\rangle_{WC,Q} =
\sum_t \alpha^t |\text{N}^t\rangle_{WC}$ ($t = \{ \; ,*,\sharp \}$).
Nucleobases A and T have two different tautomer forms, whereas G and
C have three different tautomer forms in allowed transitions
(Figures \ref{F-AT}, \ref{F-GC}). Superposition state
$|\text{N}\rangle_{WC,Q} = \alpha |\text{N}\rangle_{WC} + \alpha^*
|\text{N}^*\rangle_{WC}$ requires the entanglement of first two $WC$
edge atoms of nucleobase, whereas $|\text{N}\rangle_{WC,Q} = \alpha
|\text{N}\rangle_{WC} + \alpha^* |\text{N}^*\rangle_{WC} +
\alpha^\sharp |\text{N}^\sharp\rangle_{WC}$ requires the
entanglement of all the three $WC$ edge atoms. So, superposition of
usual and unusual tautomer forms corresponds to an intrabase
entanglement of $WC$ edge atoms of the nucleobase. Such interbase
entanglements will be invariant in the situations causing tautomeric
transitions. Thus, formation of intrabase entanglements increase the
speed of replication if they are not fragile to cellular
decoherence.

Possible quantum mechanical transitions
A$\cdot$T$\rightarrow$A$^{*}$$\cdot$T$^{*}$,
G$\cdot$C$\rightarrow$G$^{\sharp}$$\cdot$C$^{\sharp}$ and
G$\cdot$C$\rightarrow$G$^{*}$$\cdot$C$^{*}$ were found in dsDNA by
density functional theory (DFT) calculations \citep{HBPT-6,HBPT-8,
HBPT-9, HBPT-10, HBPT-11}. Hence, after base pairing, nucleobase
pairs can exist in a superposition of states corresponding to
different tautomer pairs, i.e. $|\text{N} \cdot \text{\={N}}
\rangle_{WC,O} = \sum_t \beta^t |\text{N}^t \cdot \text{\={N}}^t
\rangle_{WC}$ ($t = \{ \; ,*,\sharp \}$). However, probability
amplitudes for unusual tautomer pairs ($\beta^*$ and $\beta^\sharp$)
are expected to be very small, since transitions to unusual tautomer
pairs are very rare in dsDNA. Also, nonlocal DFT methods
\citep{HBK-1, HBK-7, HBK-2, HBK-4} showed that minimum covalent
contribution to the interbase hydrogen bonds is 38\% in A$\cdot$T
pairing and is 35\% in G$\cdot$C pairing. This can be interpreted as
quantum mechanical sharing of proton which causes an entanglement
between the donor and acceptor atoms. In this view, superposition
states $|\text{N} \cdot \text{\={N}} \rangle_{WC,O}$ produced by the
base pairing correspond to interbase entanglements.

\begin{figure}[h]
\centering
\includegraphics[width=0.48\textwidth]{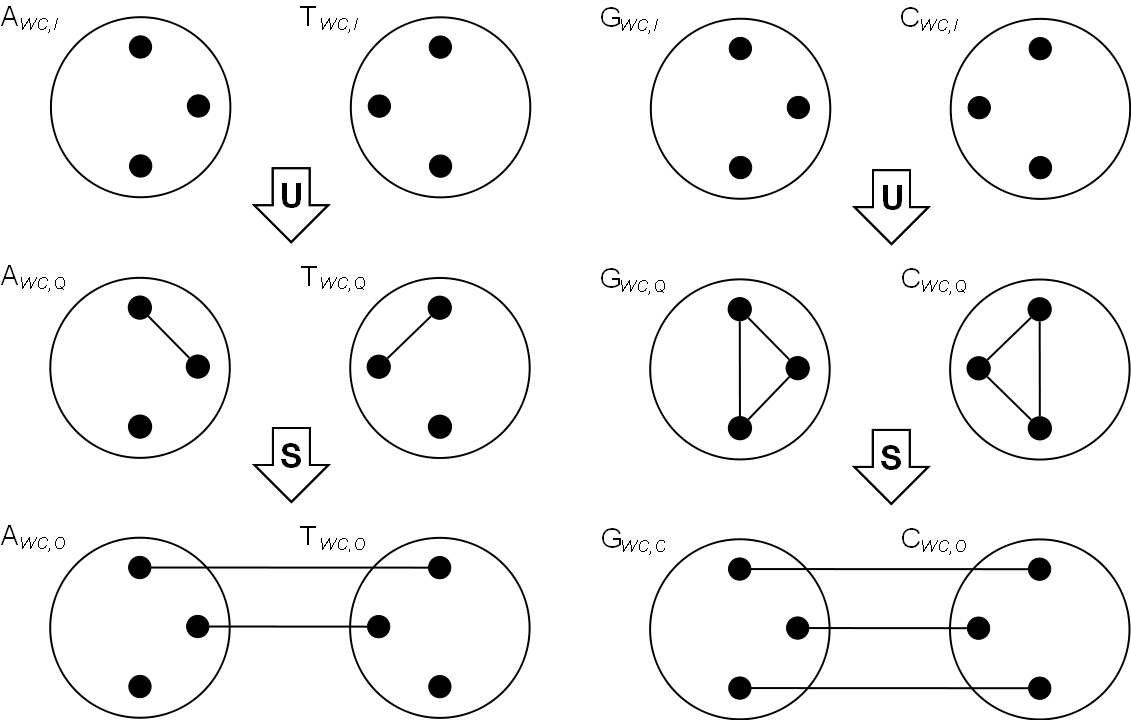}
\caption{Entanglement swapping model of replication. Each circle
represents the $WC$ edge of related nucleobase. Atoms on $WC$ edge
are shown by small dark points. A line linking two of such points
means that there is an entanglement between the atoms shown by these
points. Recognition process is described by a unitary transformation
$\mathbf{U}$: $|\text{N}\rangle_{WC,I} \rightarrow
|\text{N}\rangle_{WC,Q}$, whereas base pairing is described by an
irreversible transformation $\mathbf{S}$: $|\text{N}\rangle_{WC,Q}
\rightarrow |\text{N}\rangle_{WC,O}$} \label{F-M}
\end{figure}

We propose that recognition triggers two two-particle entanglements
in the $|\text{N}\rangle_{WC,Q}$ states of A and T and two
three-particle entanglements in the $|\text{N}\rangle_{WC,Q}$ states
of G and C (Figure \ref{F-M}). If we consider each interbase
hydrogen bond as a two-particle entanglement, there are two
two-particle entanglements in $|\text{A} \cdot
\text{T}\rangle_{WC,O}$ state and three two-particle entanglements
in $|\text{G} \cdot \text{C}\rangle_{WC,O}$ state after base
pairing. In order to turn intrabase entanglements into interbase
entanglements, recognition should be followed by an irreversible
transformation. Therefore, in our model, base pairing is described
by a multiparticle entanglement swapping \citep{KE-6} in which
DNA\emph{pol} swaps intrabase entanglements with interbase
entanglements (Figure \ref{F-M}).

While van der Waals interactions between nucleotide bases in ssDNA
were modeled by entanglement in a recent study \citep{KE-28},
present study is the first example of modeling hydrogen bonds by
entanglement and using (multiparticle) entanglement swapping in
living systems. Intuitively, biomolecules appear to be classical
objects since their de Broglie wavelengths are comparatively smaller
than their actual size due to their huge molecular mass and high
temperature. However, it is both theoretically and experimentally
shown that entanglement can occur in macroscopic and hot
non-equilibrium systems, such as biological ones \citep{KE-29,
KE-30, KE-31, KE-32}.

\subsection{Qubit Representation of Input and Output Nucleotide
States in Replication}

\begin{figure}[h]
\centering
\includegraphics[width=0.46\textwidth]{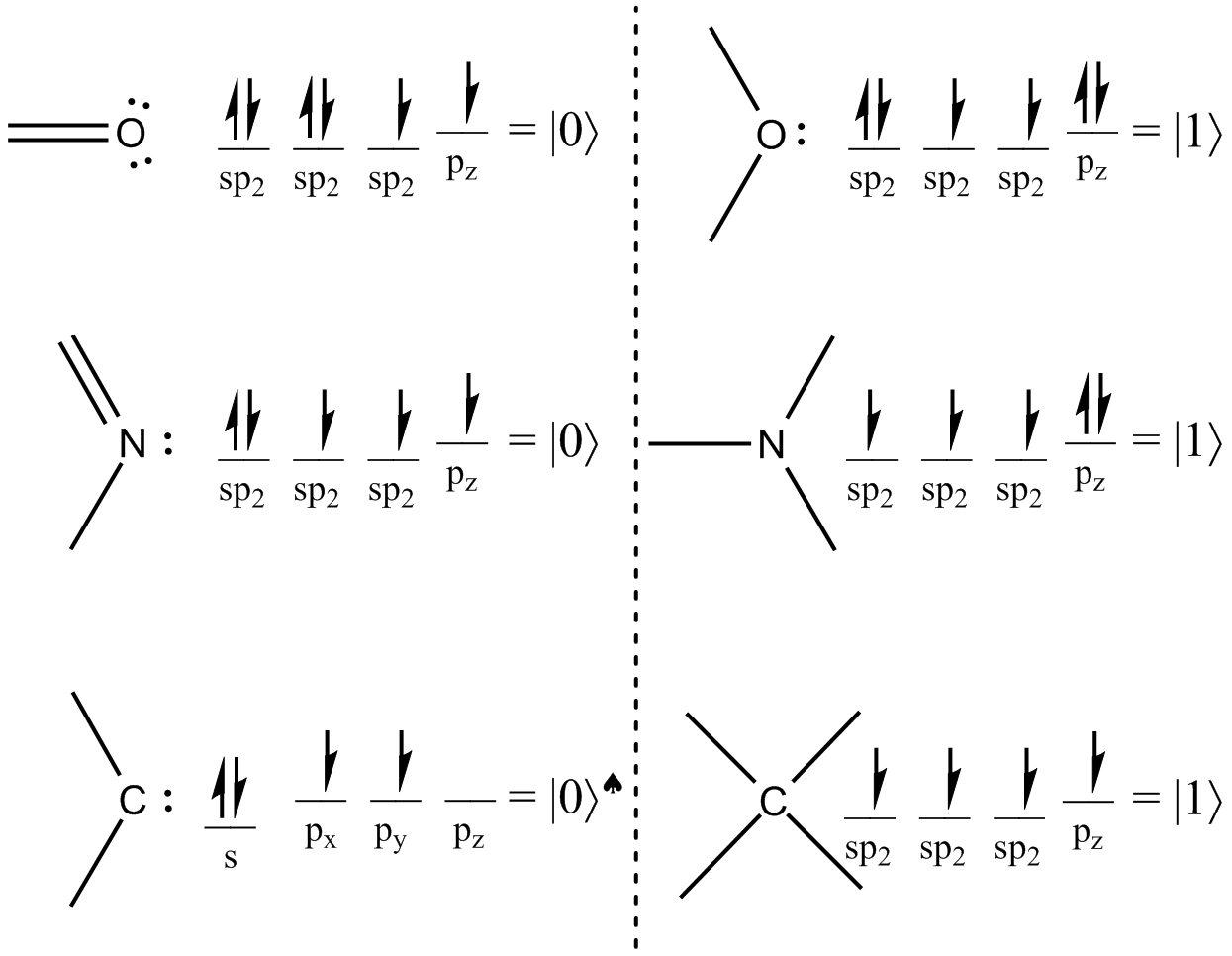}
\caption{Electronic configurations and qubit representations of the
O, N, and C atoms: higher energy state $|1\rangle$ (lower energy
state $|0\rangle$) corresponds to the presence (absence) of a proton
which is bonded to that atom and which participates in interbase
hydrogen bonds (Figures \ref{F-AT}, \ref{F-GC}). Configuration
indicated by $^{\spadesuit}$ is not present in any tautomer form.
However, it is possible to observe it in blue-shifting hydrogen
bonds of DNA.} \label{F-EC}
\end{figure}

Recognition process requires formation of at least two hydrogen
bonds between amino acid side-chains of the DNA\emph{pol} and
nucleobase \citep{KE-9, KE-8}. Such pairs of hydrogen bonds can
occur over one of the three parts of nucleotides \citep{KE-9} shown
in Figure \ref{F-parts}. In consensus, hydrogen bond donor and
acceptor atoms of bases are only O and N atoms. However, there are a
small number of computational observations in which C atoms of
nucleotide bases have the ability to make blue-shifting hydrogen
bonds \citep{HBPT-3}. In this respect, when electronic
configurations of the individual O, N, and C atoms on Hoogsteen
($H$), Watson-Crick ($WC$), and Sugar ($S$) edges (Figure
\ref{F-parts}) are considered, it is found that each atom has two
different energy states: a relatively lower energy state as acceptor
and a relatively higher energy state as donor (Figure \ref{F-EC}).
These lower and higher energy states can be regarded as qubits
$|0\rangle$ and $|1\rangle$, respectively (Figure \ref{F-EC}). Then,
reliable qubit representations can be written down for all the three
edges of each nucleobase (Table \ref{T-states}).
\begin{table}[h]
\centering \caption{Qubit representations of usual and unusual
tautomer forms found in the allowed transitions (Figures \ref{F-AT},
\ref{F-GC}): $|0\rangle$ and $|1\rangle$ states are assigned
according to the absence and presence of a proton that can be shared
in a hydrogen bond and order of the qubits are determined as shown
in Figure \ref{F-parts}.}
\begin{tabular}[c]{r @{~} l l l l}
\\
\hline Tautomer & form & $|\text{N}\rangle_H$ &
$|\text{N}\rangle_{WC}$ & $|\text{N}\rangle_S$
\\
\hline A & & $|01\rangle$ & $|101\rangle$ & $|10\rangle$
\\
A & $^{*}$ & $|00\rangle$ & $|011\rangle$ & $|10\rangle$
\\
T & & $|10\rangle$ & $|010\rangle$ & $|0\rangle$
\\
T & $^{*}$ & $|11\rangle$ & $|100\rangle$ & $|0\rangle$
\\
G & & $|00\rangle$ & $|011\rangle$ & $|10\rangle$
\\
G & $^{*}$ & $|01\rangle$ & $|101\rangle$ & $|10\rangle$
\\
G & $^{\sharp}$ & $|01\rangle$ & $|110\rangle$ & $|00\rangle$
\\
C & & $|11\rangle$ & $|100\rangle$ & $|0\rangle$
\\
C & $^{*}$ & $|10\rangle$ & $|010\rangle$ & $|0\rangle$
\\
C & $^{\sharp}$ & $|10\rangle$ & $|001\rangle$ & $|1\rangle$
\\ \hline
\end{tabular}
\label{T-states}
\end{table}

\subsection{Quantum Aspect of the Enzyme Action}

If states of the nucleobases which are measured by DNA\emph{pol}
live in a Hilbert Space whose dimension is larger than the number of
these states, there can be unavoidable efficiency problems in both
recognition and searching mechanisms. Also, states corresponding to
usual tautomer forms should be orthogonal to each other. Under these
conditions, DNA\emph{pol} should recognize bases in both free
nucleotide and ssDNA cases only over the $H$ edge according to Table
\ref{T-states}. It is known that sequence-specific dsDNA binding
proteins usually interact with the major groove \citep{KE-10} and
so, they recognize nucleobases of dsDNA over the $H$ edge, too. Such
a coincidence is not a surprise, since reading information from
dsDNA and ssDNA by different proteins ought to be based on similar
principles.

If DNA\emph{pol} makes a quantum measurement on the state
$|\text{N}\rangle_H$ to recognize a nucleobase, first qubit gives
information about purine-pyrimidine distinction, whereas the second
one gives information about imino-enol distinction. In this sense,
DNA\emph{pol} should pair bases whose qubit representations are
complementary to each other (see Table \ref{T-states}). Not only
correct base pairings, but also mispairings like A$\cdot$C$^{*}$ and
G$^{*}$$\cdot$T can be accounted for by this assumption.

A quantum measurement requires an entanglement between the measuring
device and measured system. In this case, it can be considered as a
hydrogen bonding between the DNA\emph{pol} and the nucleobase. Also,
it is reasonable to assume that entanglement between the
DNA\emph{pol} and the nucleobase should be maximal since an accurate
measurement requires strong coupling between measuring device and
measured system.
\begin{figure}[h]
\centering
\includegraphics[width=0.41\textwidth]{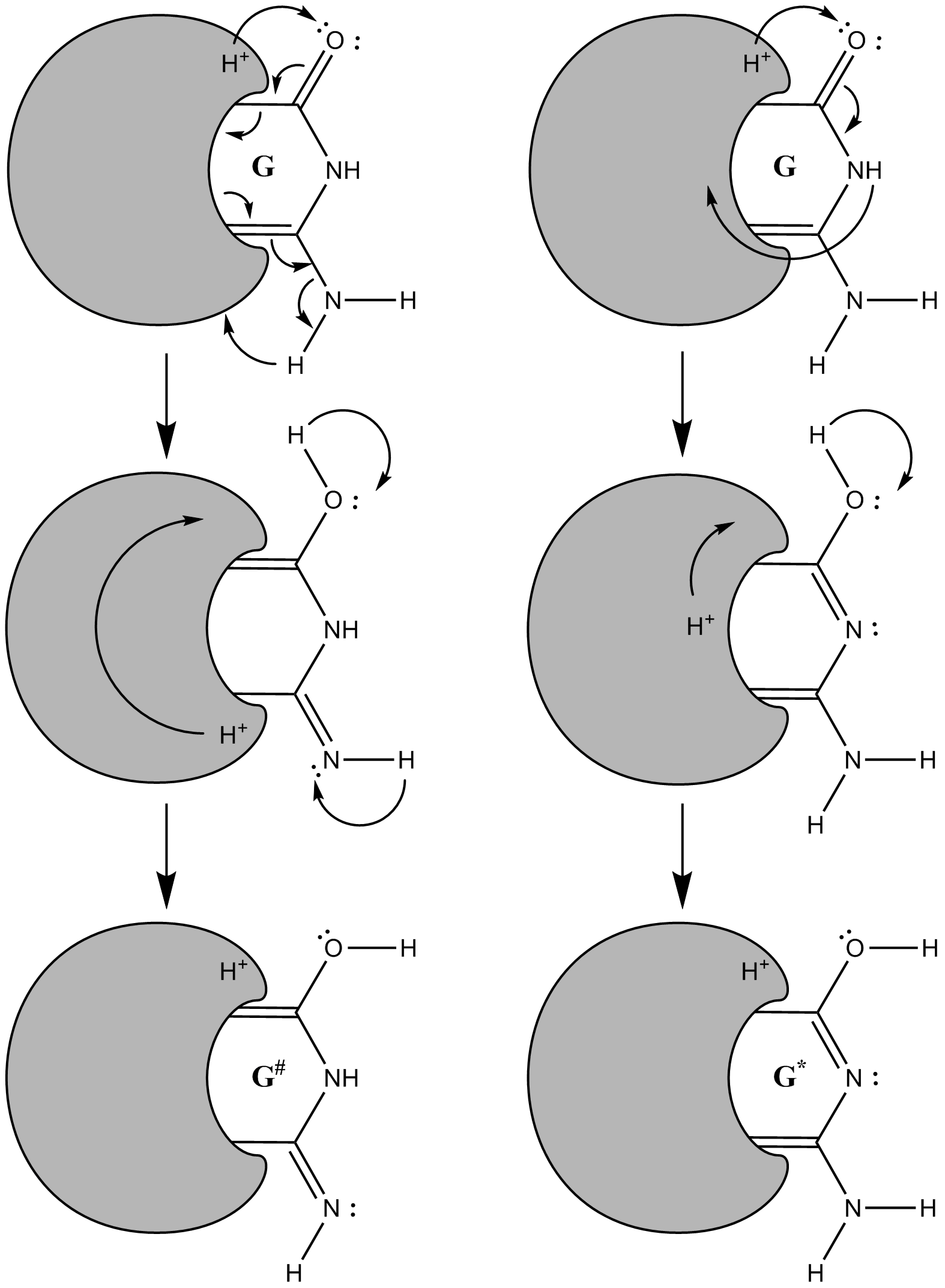}
\caption{An hypothetical mechanism for the tautomeric transitions of
nucleobase G by proton transfer between enzyme DNA\emph{pol} and
nucleotide. Grey structure represents the active site of
DNA\emph{pol} and arrows show proton and electron delocalizations.}
\label{F-E}
\end{figure}

Hypothetically, a proton transfer between the DNA\emph{pol} and the
second atom of $H$ edge (or equally the first atom of $WC$ edge)
which occurs during the recognition, can trigger a tautomeric
transition (Figure \ref{F-E}). Since such a transfer has a quantum
nature in a hydrogen bonding, recognition can trigger a transition
to the superposition of usual and unusual tautomer forms by a
unitary transformation $\mathbf{U}$. This mechanism is a toy model
of evolution of the basis states $|\text{N}\rangle_{WC,I}$ (Table
\ref{T-states}) into superposition states $|\text{N}\rangle_{WC,Q} =
\sum_t \alpha^t |\text{N}^t\rangle_{WC,I}$ ($t = \{ \; ,*,\sharp
\}$) as follows:
\begin{eqnarray}
\label{E-Q} |\text{A}\rangle_{WC,I}
\overset{\text{U}}{\longrightarrow} |\text{A}\rangle_{WC,Q} &=&
a|101\rangle + a^{*}|011\rangle \, ,
\\
|\text{T}\rangle_{WC,I} \overset{\text{U}}{\longrightarrow}
|\text{T}\rangle_{WC,Q} &=& t|010\rangle + t^{*}|100\rangle \, ,
\nonumber
\\
|\text{G}\rangle_{WC,I} \overset{\text{U}}{\longrightarrow}
|\text{G}\rangle_{WC,Q} &=& g |011\rangle + g^{*}|101\rangle +
g^{\sharp}|110\rangle \, , \nonumber
\\
|\text{C}\rangle_{WC,I} \overset{\text{U}}{\longrightarrow}
|\text{C}\rangle_{WC,Q} &=& c |100\rangle + c^{*}|010\rangle +
c^{\sharp}|001\rangle \, . \nonumber
\end{eqnarray}

Now, it is clearly seen that these superpositions of different
tautomer forms are nothing else than intrabase entanglements of the
atoms on $WC$ edge. Nucleobases A and T have two different tautomer
forms, whereas G and C have three different tautomer forms in
allowed transitions (Figures \ref{F-AT}, \ref{F-GC}). Thus, after
recognition, we observe two two-qubit entanglements in the states of
A and T, while there are two three-qubit entanglements in the states
of G and C (Figure \ref{F-M}).

Recognition process of complementary nucleobase \={N} also involves
a quantum measurement in which a maximal entanglement is formed
between DNA\emph{pol} and \={N}. However, DNA\emph{pol} can not bind
to \={N} in such a quantum mechanical way until it disentangles
itself from N. This is because of the entanglement monogamy (or
polygamy) \citep{monogamy_1, monogamy_2, monogamy_3} which roughly
says that if A and B are maximally entangled, then any one of them
can not be simultaneously entangled with C. In the context of
monogamy, formation of intrabase entanglement in N breaks the
maximal entanglement between DNA\emph{pol} and N. Then, a maximal
entanglement between DNA\emph{pol} and \={N} becomes possible.

Similarly, recognition of \={N} induces an intrabase entanglement in
\={N} which disentangles DNA\emph{pol} from \={N}. This
disentanglement allows DNA\emph{pol} to bond N and \={N} together
and then to bind to the subsequent N of ssDNA in a quantum
mechanical way. Therefore, formation of intrabase entanglements not
only prevents the uncontrollable tautomeric transitions caused by
cellular environment, but also provides separation of DNA\emph{pol}
from one nucleotide and binding of it to another.

After base pairing, nucleobase pairs should exist in a superposition
of states corresponding to different tautomer pairs. Since state of
a hydrogen bonded atom pair can be written as the Bell state
$|\beta_{01}\rangle = (|01\rangle + |10\rangle) / \sqrt{2}$, these
superposition states $|\text{N}_1\cdot \text{\={N}}_2\rangle_{WC,O}$
are actually interbase entanglements. Therefore, it can be said that
in the case of G$\cdot$C pair, there are three two-qubit
entanglements in $|\beta_{01}\rangle$ state, and in the case of
A$\cdot$T pair, there are two two-qubit entanglements in
$|\beta_{01}\rangle$ state. In order to turn intrabase entanglements
into interbase entanglements, $\mathbf{U}$ should be followed by an
irreversible transformation $\mathbf{S}$. Thus, in our model, base
pairing occurs as a multiparticle entanglement swapping in which
DNA\emph{pol} swaps intrabase entanglements with interbase
entanglements (Figure \ref{F-M}).

\subsection{Quantum Circuit for Intrabase Entanglement}
\begin{figure}[h]
\centering
\includegraphics[width=0.47\textwidth]{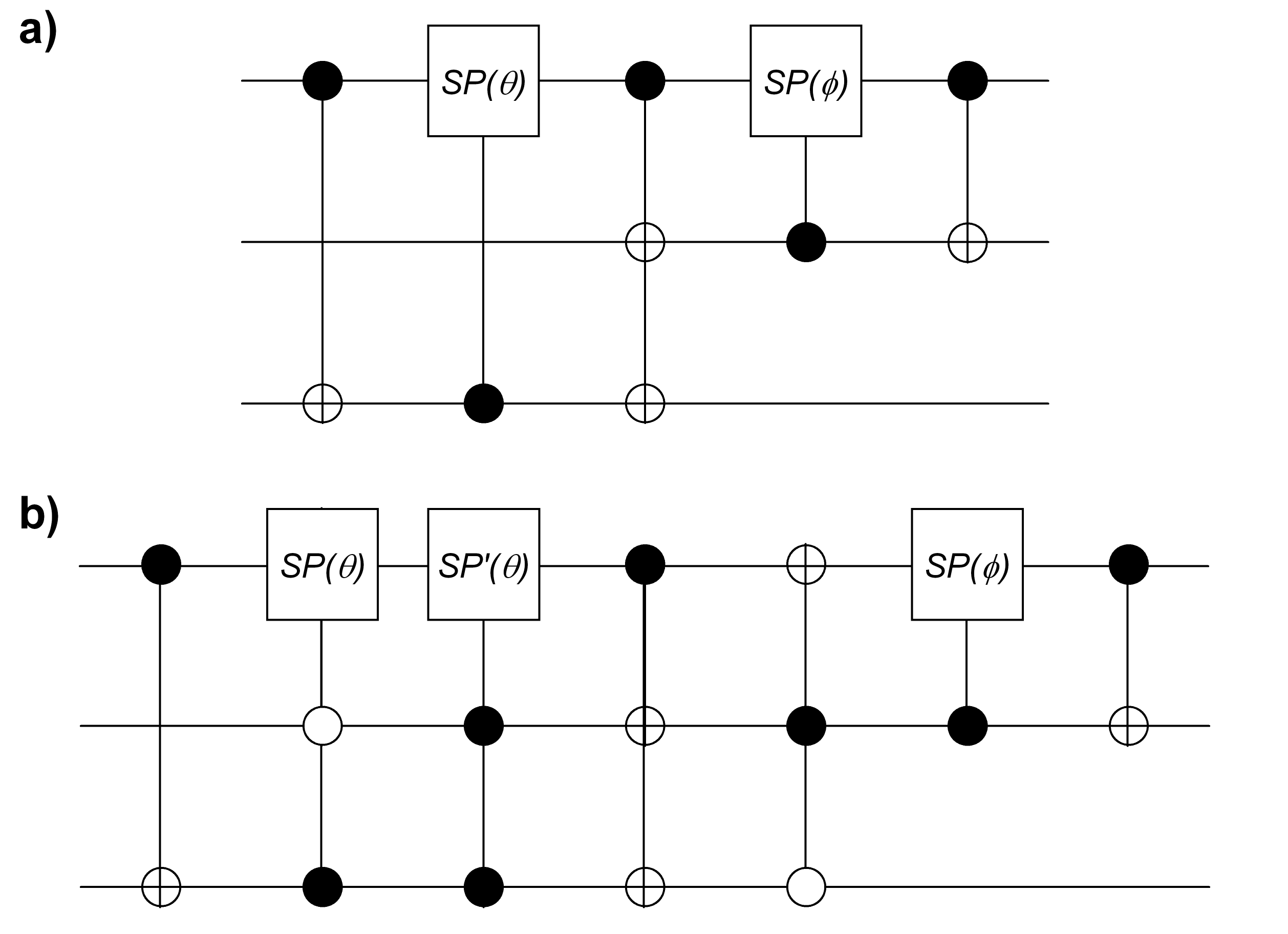}
\caption{Two of the possible quantum circuits for transformation
$\mathbf{U}$ which turns $|\text{N}\rangle_{WC,I}$ states into the
$|\text{N}\rangle_{WC,Q}$ states. Superposition matrices
$S\!P(\theta)$ and $S\!P'(\theta)$ of $controlled-S\!uperposition$
gates equals to the multiplication of rotation matrix $R(\theta)$
and Pauli-$Z$ matrix with different orders: $S\!P(\theta) =
R(\theta) \times Z$ and $S\!P'(\theta) = Z \times R(\theta)$.}
\label{F-QC}
\end{figure}

Two candidates for the transformation $\mathbf{U}$ are shown in the
Figure \ref{F-QC}. To provide a decoherence-free (DF) subsystem
\citep{KE-20, KE-21, KE-36, KE-19}, we take the angles $\theta$ and
$\phi$ in the second quantum circuit (Figure \ref{F-QC}-b) as
$\arccos(\sqrt{1} / \sqrt{3})$ and $\arccos(1 / \sqrt{2})$,
respectively. Then, $|\text{N}\rangle_{WC,Q}$ states are obtained
as:
\begin{eqnarray} \label{E-1}
|\text{A}\rangle_{WC,Q} \! \! \! \! &=& \! \! \! \! (+ \,
|011\rangle-|101\rangle)/\sqrt{2} \, ,
\\
|\text{T}\rangle_{WC,Q} \! \! \! \! &=& \! \! \! \! (+ \,
|010\rangle-|100\rangle)/\sqrt{2} \, , \nonumber
\\
|\text{G}\rangle_{WC,Q} \! \! \! \! &=& \! \! \! \! (+ \,
|011\rangle+|101\rangle-2|110\rangle)/\sqrt{6} \, , \nonumber
\\
|\text{C}\rangle_{WC,Q} \! \! \! \! &=& \! \! \! \! (- \,
|100\rangle-|010\rangle+2|001\rangle)/\sqrt{6} \, . \nonumber
\end{eqnarray}

To consider each base pair as an intact system, tensor products of
these states should be taken.
\begin{eqnarray} \label{E-2}
|\text{A} \otimes \text{T}\rangle_{WC,Q} =
\frac{1}{2}(|011\rangle|010\rangle \! \! \! \! \! \! &-& \! \! \! \!
\! \! |011\rangle|100\rangle
\\
\frac{}{}- |101\rangle|010\rangle \! \! \! \! \! \! &+& \! \! \! \!
\! \! |101\rangle|100\rangle) \, , \nonumber
\\
|\text{G} \otimes \text{C}\rangle_{WC,Q} =
\frac{-2}{\sqrt{3}}(\frac{|011\rangle|100\rangle}{4} \! \! \! \! \!
\! &+& \! \! \! \! \! \! \frac{|101\rangle|100\rangle}{4} \! - \!
\frac{|110\rangle|100\rangle}{2} \nonumber
\\ + \frac{|011\rangle|010\rangle}{4} \! \! \! \! \! \! &+&
\! \! \! \! \! \! \frac{|101\rangle|010\rangle}{4} \!  - \!
\frac{|110\rangle|010\rangle}{2} \nonumber
\\ - \frac{|011\rangle|001\rangle}{2} \! \! \! \! \! \! &-&
\! \! \! \! \! \! \frac{|101\rangle|001\rangle}{2} \!  + \!
|110\rangle|001\rangle) \, . \nonumber
\end{eqnarray}

We reorder qubits of these product states in such a way that
hydrogen bonded atom pairs come next to each other in order to
clarify base pairing. Then, we get
\begin{eqnarray} \label{E-3}
|\text{A} \cdot \text{T}\rangle_{WC,Q} =
\frac{1}{2}(|00\rangle|11\rangle|10\rangle \! \! \! \! \! \! &-& \!
\! \! \! \! \! |01\rangle|10\rangle|10\rangle
\\ \frac{}{}- |10\rangle|01\rangle|10\rangle \! \! \! \! \! \! &+& \! \! \! \! \! \!
|11\rangle|00\rangle|10\rangle) \, , \nonumber
\\
|\text{G} \cdot \text{C}\rangle_{WC,Q} =
\frac{-2}{\sqrt{3}}(\frac{|01\rangle|10\rangle|10\rangle}{4} \! \!
\! \! \! \! &+& \! \! \! \! \! \!
\frac{|11\rangle|00\rangle|10\rangle}{4} \! - \!
\frac{|11\rangle|10\rangle|00\rangle}{2} \nonumber
\\ + \frac{|00\rangle|11\rangle|10\rangle}{4} \! \! \! \! \! \! &+& \! \! \! \! \! \! \frac{|10\rangle|01\rangle|10\rangle}{4} \!  - \! \frac{|10\rangle|11\rangle|00\rangle}{2} \nonumber
\\ - \frac{|00\rangle|10\rangle|11\rangle}{2} \! \! \! \! \! \! &-& \! \! \! \! \! \! \frac{|10\rangle|00\rangle|11\rangle}{2} \!  + \!  |10\rangle|10\rangle|01\rangle) \, . \nonumber
\end{eqnarray}

\subsection{Swapping Protocol for Interbase Entanglement}
\begin{figure}[h]
\centering
\includegraphics[width=0.45\textwidth]{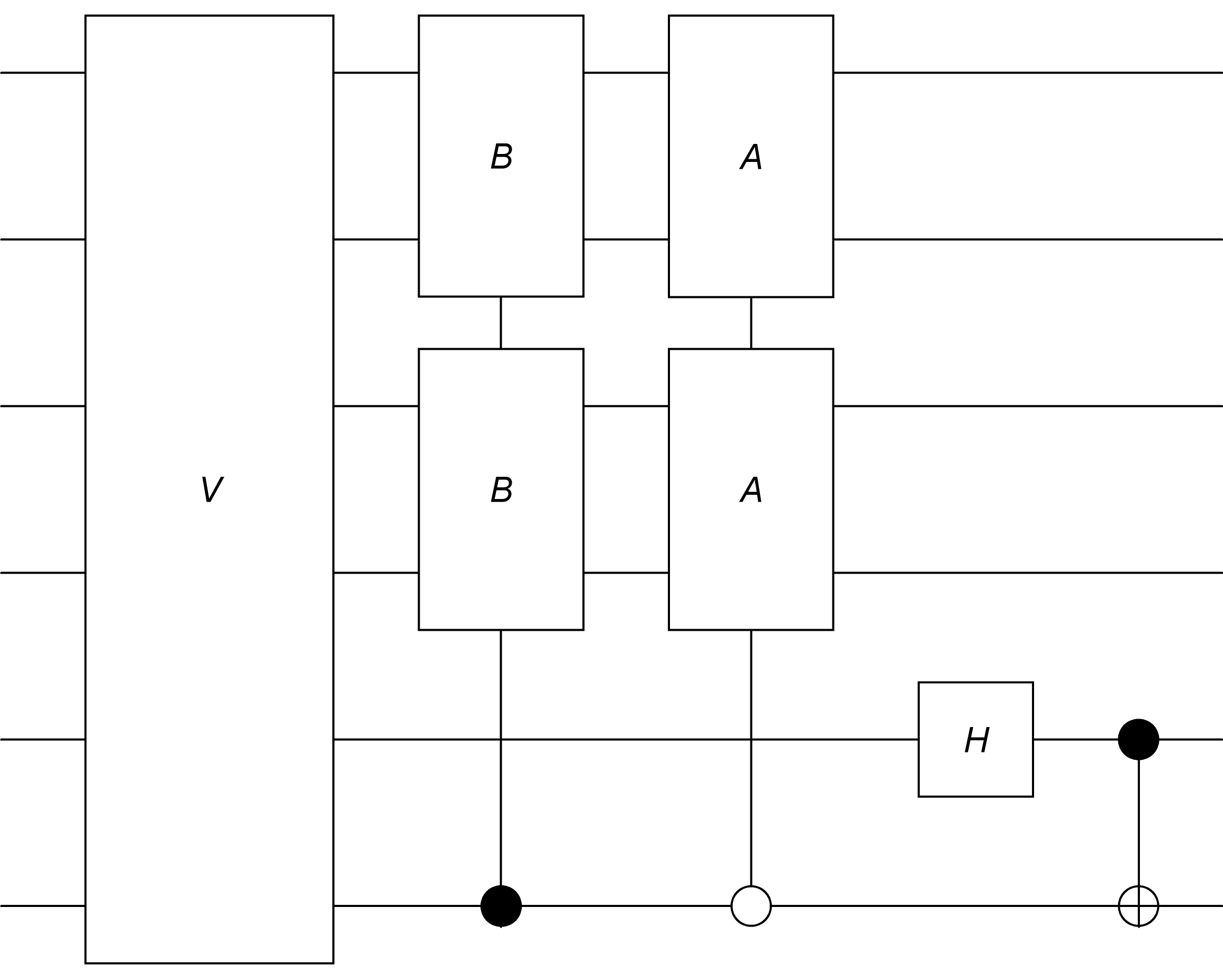}
\caption{Quantum circuit for $\mathbf{S}$ which swaps intrabase
entanglements to interbase entanglements: $H$ is the $Hadamard$ gate
which equals to $S\!P(\pi/4)$ defined in Figure \ref{F-QC}.
Transformations $A$ and $B$ are shown in the subsequent figure. See
Appendix for the details of transformation $V$.} \label{F-QC2}
\end{figure}

Swapping intrabase entanglements to interbase entanglements can be
achieved by a three-step protocol $\mathbf{S}$ as follows:

1. Reordered base pair states are subjected to a transformation $V$
as shown in Figure \ref{F-QC4}. Then, fifth and sixth qubits of the
G$\cdot$C (or C$\cdot$G) pair become $|01\rangle$, whereas fifth and
sixth qubits of the A$\cdot$T (or T$\cdot$A) pair become
$|11\rangle$. After this transformation, any improper base pair
exists in a superposition of states in which third qubit pair is
always $|00\rangle$ or $|10\rangle$.

\begin{figure}[h]
\centering
\includegraphics[width=0.41\textwidth]{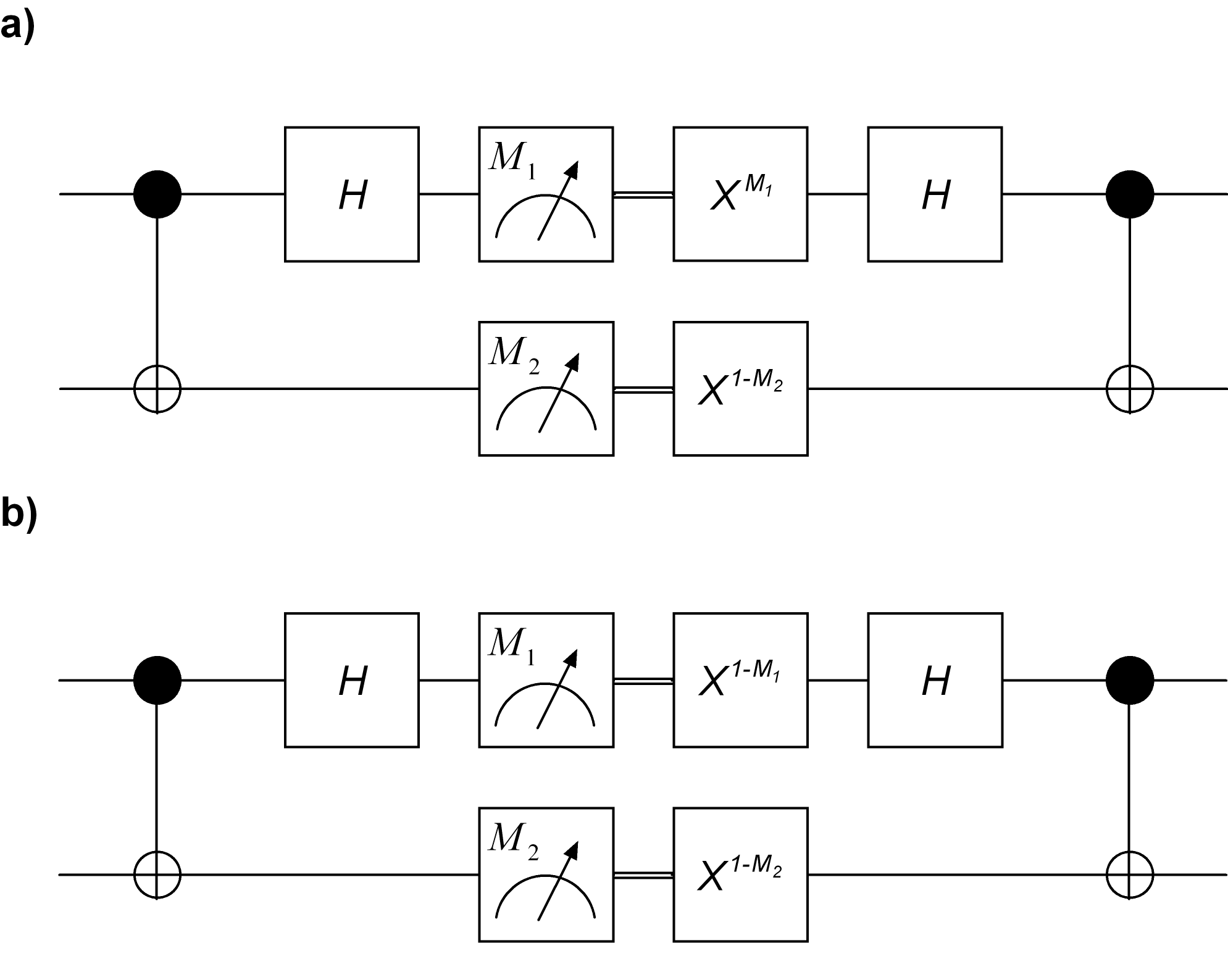}
\caption{Quantum transformations $A$ and $B$ used in $\mathbf{S}$:
they are actually modified Bell measurements. Modifications made by
$X$ which is the $NOT$ gate (or Pauli-$X$ gate) and $X$'s
superscript ($M_1$ or $M_2$) is the outcome of the measurement which
is done immediately before it. \emph{a}) $A$ transforms any Bell
state $|\beta_{ij}\rangle$ into the Bell state $|\beta_{11}\rangle$.
\emph{b}) $B$ transforms any Bell state $|\beta_{ij}\rangle$ into
the Bell state $|\beta_{01}\rangle$.} \label{F-QC3}
\end{figure}

2. If the sixth qubit is $|0\rangle$, first and second qubit pairs
undergo a transformation $A$ as shown in Figure \ref{F-QC3}-a.
Otherwise, these qubit pairs are transformed with transformation $B$
(Figure \ref{F-QC3}-b). Then, first and second qubit pairs of proper
base pairs collapse into Bell state $|\beta_{01}\rangle =
(|01\rangle + |10\rangle)/\sqrt{2}$, whereas first and second qubit
pairs of improper base pairs collapse into Bell state
$|\beta_{11}\rangle = (|01\rangle - |10\rangle)/\sqrt{2}$.

3. Firstly, fifth qubit is passed through a $Hadamard$ ($H$) gate.
Then, sixth qubit is converted by $NOT$ ($X$) gate if the fifth
qubit is $|1\rangle$. After this step, third qubit pair of the
G$\cdot$C (or C$\cdot$G) pair becomes $|\beta_{01}\rangle$, whereas
third qubit pair of the A$\cdot$T (or T$\cdot$A) pair becomes
$|\beta_{11}\rangle$. In contrast, fifth and sixth qubits of any
improper base pair exists in one of the Bell states
$|\beta_{00}\rangle = (|00\rangle + |11\rangle)/\sqrt{2}$ or
$|\beta_{10}\rangle = (|00\rangle - |11\rangle)/\sqrt{2}$.

Immediately after $\mathbf{S}$, proper $|\text{N}_1\cdot
\text{\={N}}_2\rangle_{WC,O}$ states are written in terms of the
Bell states as follows.
\begin{eqnarray} \label{E-4}
|\text{A}\cdot\text{T}\rangle_{WC,O} \! \! \!\!&=&\! \! \!\!
|\text{T}\cdot\text{A}\rangle_{WC,O} =
|\beta_{01}\rangle|\beta_{01}\rangle|\beta_{11}\rangle ,
\\
|\text{G}\cdot\text{C}\rangle_{WC,O} \! \! \!\!&=&\! \! \!\!
|\text{C}\cdot\text{G}\rangle_{WC,O} =
|\beta_{01}\rangle|\beta_{01}\rangle|\beta_{01}\rangle . \nonumber
\end{eqnarray}

\subsection{Biochemical Realizations of Quantum Circuits/Protocols}

Pauli-$X$ transformation of controlled-\emph{NOT} gates used in the
quantum circuit of $\mathbf{U}$ (Figure \ref{F-QC}-b) converts
$|1\rangle_{N}$ into $|0\rangle_{N}$. When state of the
DNA\emph{pol} is also considered with the subscript $E$, this
transformation should be $|1\rangle_{N}|0\rangle_{E} \rightarrow
|0\rangle_{N}|1\rangle_{E}$. Since $|0\rangle$ and $|1\rangle$
states of an atom respectively correspond to the absence and
presence of a proton bonded to that atom, this transformation can be
regarded as a proton tunneling from the nucleobase to DNA\emph{pol}
through the atom on which the gate acts. Vice versa is possible for
the action of Pauli-$X$ transformation on the state $|0\rangle_{N}$.

Other gates used in the quantum circuit of $\mathbf{U}$ are
controlled-\emph{SP} and -\emph{SP$'$} gates. When argument of
\emph{SP} transformation equals to $\arccos(1 / \sqrt{2})$, it
transforms $|0\rangle_{N}$ into $(|0\rangle + |1\rangle)/\sqrt{2}$
and $|1\rangle_{N}$ into $(|0\rangle - |1\rangle)/\sqrt{2}$. The
former transformation should be $|0\rangle_{N}|1\rangle_{E}
\rightarrow |\beta_{01}\rangle_{NE}$ by taking into account also the
state of DNA\emph{pol}, whereas the latter transformation should be
$|1\rangle_{N}|0\rangle_{E} \rightarrow |\beta_{11}\rangle_{NE}$.
So, the action of the \emph{SP} transformation on the state
$|0\rangle_{N}$ can be considered as formation of a hydrogen bond
between the nucleobases and enzyme through the atom on which gate
acts. On the contrary, an antibonding should occur by the action of
\emph{SP} transformation on the state $|1\rangle_{N}$. This is
because of the fact that free energy in the state
$|\beta_{11}\rangle_{NE}$ is greater than the one in which there is
no interaction. Since entanglement measure of the generated state
changes when the argument is changed, action of \emph{SP}
transformation can produce bondings/antibondings with different
strengths for different arguments. Action of \emph{SP$'$}
transformation on $|0\rangle_{N}$ produces the same state as the
\emph{SP} action on $|1\rangle_{N}$ and vice versa. Thus, action of
\emph{SP$'$} transformation can be also considered as
bondings/antibondings.
\begin{figure}[h]
\centering
\includegraphics[width=0.45\textwidth]{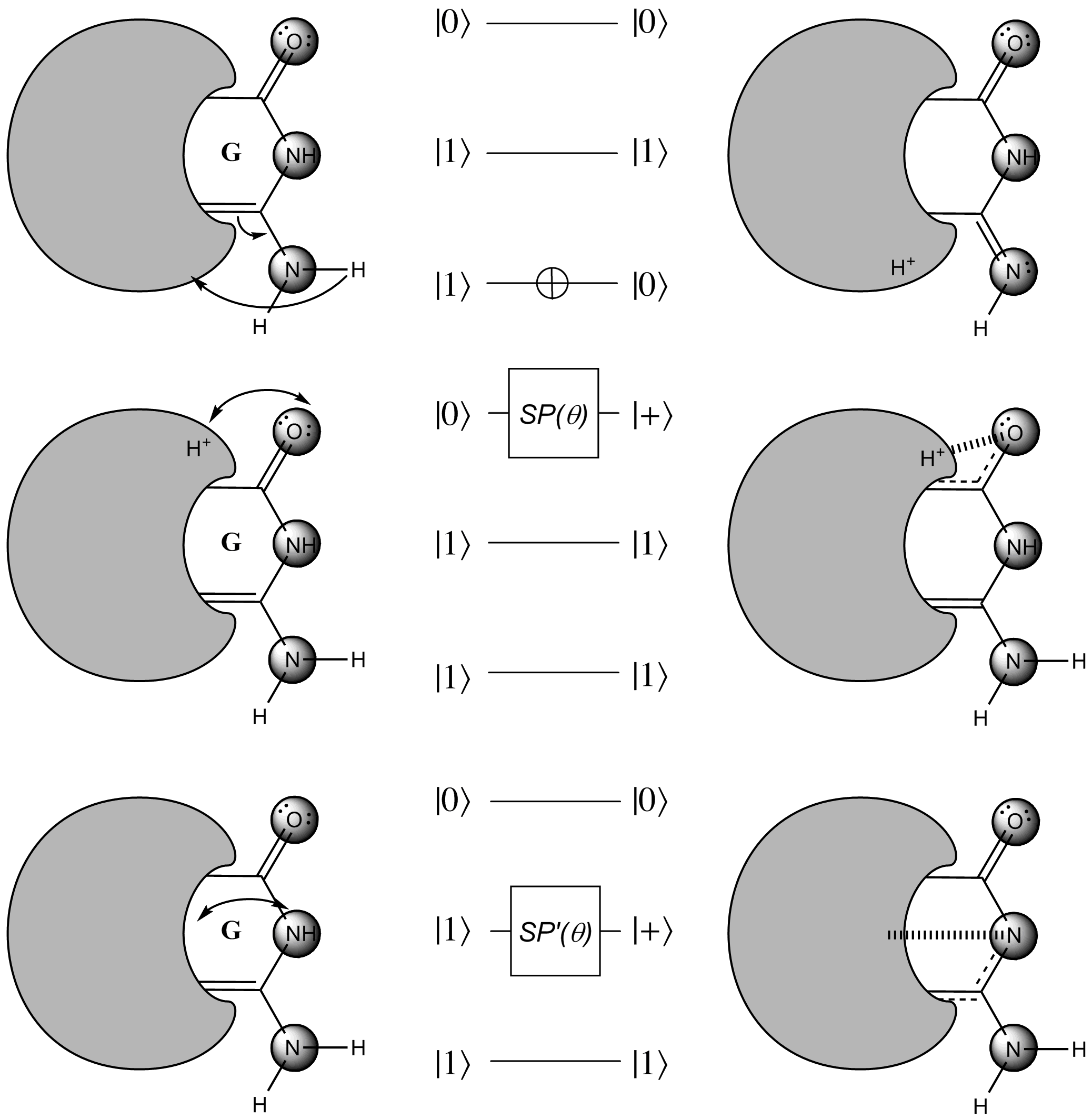}
\caption{Simple depictions of the actions of the \emph{NOT},
\emph{SP}, and \emph{SP$'$} gates: atoms whose energy states are
represented by qubits are the ones inside the shaded spheres.
Evolution of the $|N\rangle_{WC}$ state of the nucleobase due to the
action of gate is shown from left to right and $|+\rangle$ qubit
equals to $(|0\rangle + |1\rangle)/\sqrt{2}$ if $\theta$ equals to
$\arccos(1 / \sqrt{2})$.} \label{F-dep1}
\end{figure}
\begin{figure}[h]
\centering
\includegraphics[width=0.49\textwidth]{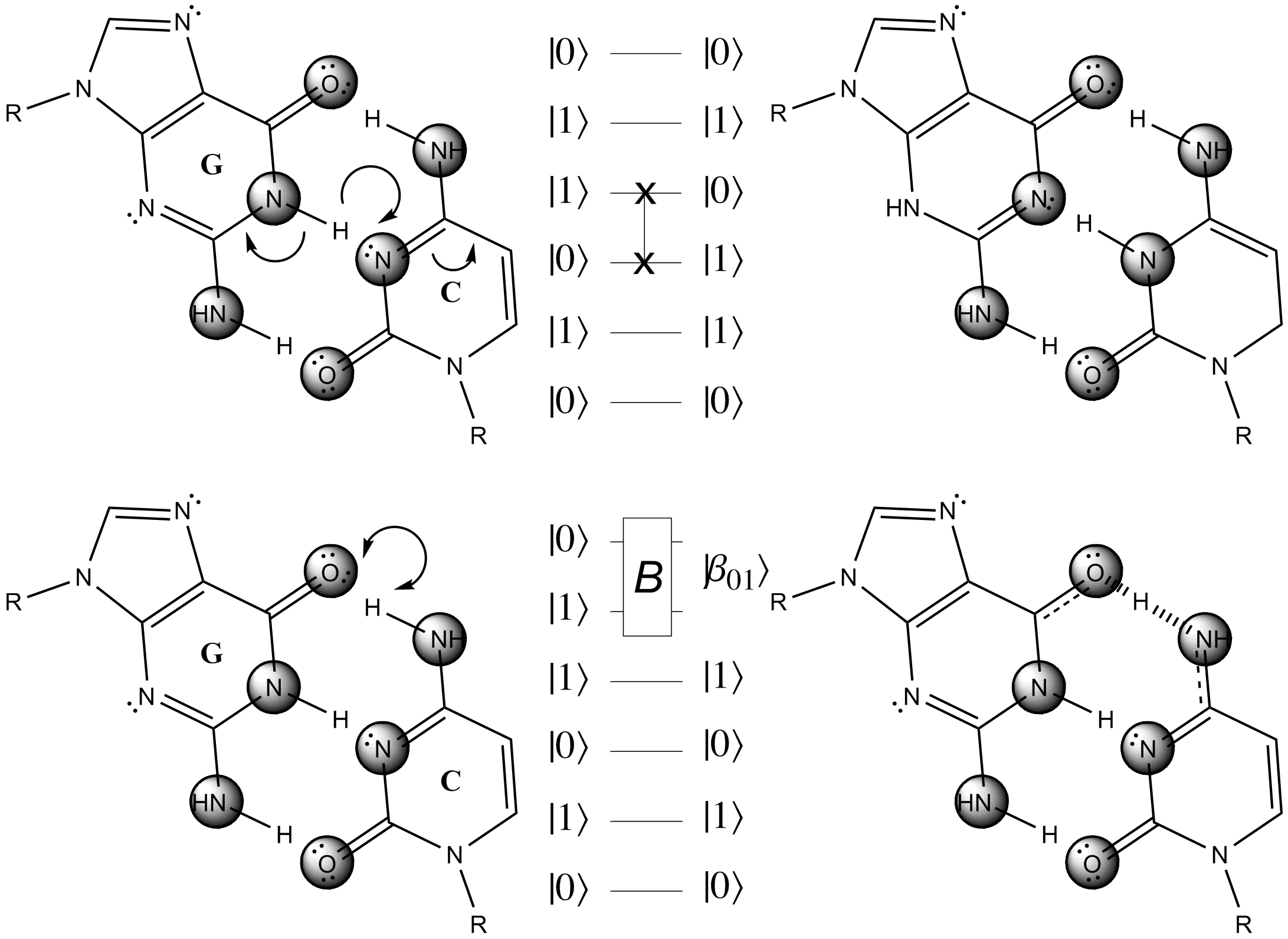}
\caption{Simple depictions of the actions of the swap gate and
modified Bell measurement $B$: atoms whose energy states are
represented by qubits are the ones inside the shaded spheres.
Evolution of the reordered $|\text{N}\cdot \text{\={N}}\rangle_{WC}$
state of the nucleobase pair due to the action of gate is shown from
left to right and $|\beta_{01}\rangle$ Bell state equals to
$(|01\rangle + |10\rangle)/\sqrt{2}$.} \label{F-dep2}
\end{figure}

Both the proton transfer and hydrogen bonding are the usual tasks
done by enzymes and there are some evidences for the unignorable
role of quantum effects and dynamics on the enzymatic reactions
\citep{KE-11, KE-12, KE-13}. Therefore producing an intrabase
entanglement by transformation $\mathbf{U}$ is a possible action
performed by the enzyme DNA\emph{pol} (Figure \ref{F-dep1}).

Besides $\mathbf{U}$, swapping protocol $\mathbf{S}$ also includes
controlled-\emph{NOT}, -\emph{SP}, and -\emph{SP$'$} gates which are
regarded as respectively proton tunneling and hydrogen
bonding/antibonding between a nucleobase and DNA\emph{pol}.
Additionally, $\mathbf{S}$ contains swap gates (Figure \ref{F-QC4})
and modified Bell measurements ($A$ and $B$). Swap gate exchanges
the states of two qubits on which it acts: $|10\rangle \rightarrow
|01\rangle$ and $|01\rangle \rightarrow |10\rangle$. So, it can be
interpreted as a proton tunneling similar to interpretation of
Pauli-$X$ transformation of controlled-\emph{NOT} gate. However,
atoms on which swap gate acts can belong either to the same
nucleobase or to the different nucleobases in the base pair. Hence,
this proton tunneling should be considered inside a nucleobase or
between the two different bases.

Outcome of the modified Bell measurement $B$ is the Bell state
$|\beta_{01}\rangle$, whereas the outcome of the modified Bell
measurement $A$ is the Bell state $|\beta_{11}\rangle$. Thus, these
measurements can also be thought as a hydrogen bonding/antibonding.
Contrary to the ones produced by \emph{SP} and \emph{SP$'$} gates,
this bonding/antibonding is between the two nucleobases and its
strength is always maximum. Consequently, $\mathbf{S}$ consists of
nothing more than proton tunneling and hydrogen
bondings/antibondings which are the usual tasks done by enzymes like
DNA\emph{pol} (Figures \ref{F-dep1}, \ref{F-dep2}).

Immediately after the entanglement swapping, states of proper base
pairs are found as in Equation \ref{E-4}. According to these states,
A$\cdot$T (or T$\cdot$A) base pair has two hydrogen bonds and
G$\cdot$C (or C$\cdot$G) base pair has three hydrogen bonds as in
the actual case. However, these hydrogen bonds have a maximum
strength since Bell states are maximally entangled. Amplitudes in
these states should change by the quantum evolution in the presence
of the asymmetric double well potentials of the hydrogen bonded atom
pairs. Thus, strengths of the hydrogen bonds should gradually
decrease to the actual ones. Moreover, there is an antibonding
between the last atom pair in A$\cdot$T (or T$\cdot$A). These atoms
repulse each other because of the higher free energy of antibonding,
but strength of this repulsion should also decrease by time. Since
one of the atoms in this antibonding is C atom, final strength of
the repulsion should be negligible.

On the other hand, both first and second atom pairs of the improper
base pairs have an antibonding after the entanglement swapping.
Final strength of the repulsions due to these antibonding
interactions are not negligible and so, they should destabilize and
separate the improper base pairs. However, state of the last atom
pair in these base pairs are obtained as $|\beta_{00}\rangle$ or
$|\beta_{10}\rangle$. Since total proton number of the base pair
does not remain constant after collapsing to these states, atom pair
and DNA\emph{pol} can not separate from each other. In fact, these
Bell states should be treated as an entanglement between the atom
pair and DNA\emph{pol} when state of the enzyme is also under
consideration: $|00\rangle_{N \emph{\text{\={N}}}}\pm|11\rangle_{N
\emph{\text{\={N}}}} \rightarrow |00\rangle_{N
\emph{\text{\={N}}}}|11\rangle_{E}\pm|11\rangle_{N
\emph{\text{\={N}}}}|00\rangle_{E}$. It can be that it is this
entanglement which keeps DNA\emph{pol} in place till the correct
\={N} comes along. Both of the asymmetric potentials and
destabilization of the base pair should weaken this entanglement.
When entanglement is weakened enough, DNA\emph{pol} can bind to the
correct \={N} because of the converse monogamy \citep{monogamy_4}
which roughly says that if A and B are weakly entangled, then any
one of them could be strongly entangled with C. After that, a
Pauli-$X$ transformations can fix the total number of protons on the
improper base pair and make incorrect \={N} separable from the
complex.

Neither $\mathbf{U}$ nor $\mathbf{S}$ is unique for the given model.
However, this is not a disadvantage since there are several
DNA\emph{pol} species and families with different replication
fidelities. This diversity in replication fidelity of DNA\emph{pol}
can be accomplished by different $\mathbf{U}$ and $\mathbf{S}$
pairs.

\subsection{Effects of Cellular Decoherence}

The intact system which is exposed to decoherence is the whole
nucleobase - DNA\emph{pol} complex. Hence, states of the nucleobases
alone are not sufficient to determine if decoherence has a
significant effect on the transformation $\mathbf{U}$ or does not.
To draw a complete picture of interaction, assume that there are $q$
hydrogen bond acceptors and $(k-q)$ hydrogen bond donors in the
active site of DNA\emph{pol}. If so, enzyme's active site can be
represented by the state $|0\rangle^{\otimes
q}_{E}\otimes|1\rangle^{\otimes (k-q)}_{E}$ after a proper ordering
in which all $|0\rangle$ qubits are put to left of all $|1\rangle$
qubits. Then, we can obtain the initial state of the nucleobase -
DNA\emph{pol} complex as $|s\rangle_I =
|\text{N}\rangle_{WC,I}\otimes|0\rangle^{\otimes
q}_{E}\otimes|1\rangle^{\otimes (k-q)}_{E}$. Cellular decoherence
effect on the state $|s\rangle$ can be simplified as a weak
collective decoherence \citep{KE-36, KE-19} which turns $|1\rangle$
states into $e^{i\phi}|1\rangle$, while $|0\rangle$ states remain
unchanged. Since we have already considered $|0\rangle$ and
$|1\rangle$ states of an atom respectively as the absence and
presence of a proton bonded to that atom, this simplification makes
sense: decoherence can not affect an absent proton.

When spacing between the qubits is smaller than the wavelength of
the radiation field which acts as a boson bath for the qubit system,
collective decoherence dominates among the others \citep{KE-20,
KE-36, KE-19}. In this sense, weak collective decoherence seems to
be relevant not only to long-range electrostatic interactions with
the intracellular ions, but also to short-living couplings with the
thermal reservoir. Investigation of the effect of thermal reservoir
on the internal DNA mobility requires a lattice dynamic approach
based on an atomistic description of the molecule \citep{KE-33}.
According to the appropriate methods given in \citep{KE-34, KE-35},
maximum frequency of the vibrational modes in DNA is a few hundreds
of cm$^{-1}$ at room temperature. Corresponding phonon wavelength is
in the order of $\mu$m and this is quite longer than the qubit
spacing in our model, which is no more than $3 \, {\AA}$. Thus,
phonon bath can not distinguish the qubits and collective
decoherence is expected to be the dominant decoherence mechanism in
the DNA replication.

Defining a variable $\lambda_K$ which equals to the number of
$|0\rangle$ qubits minus the number of $|1\rangle$ qubits in a state
over $K$ qubits, \cite{KE-36, KE-19} showed that subspaces of
Hilbert space spanned by the states with constant $\lambda_K$ are
decoherence-free (DF) during a collective dephasing process as
described above. Then, a DF subspace for a specific $\lambda_K$ is
denoted as $DFS_{K}(\lambda_K)$ \citep{KE-36, KE-19}.

In our model, there is no proton exchange between the system and its
environment during the transformation $\mathbf{U}$. For example,
when a $|1\rangle$ qubit of nucleobase turns into $|0\rangle$ qubit
after a controlled-\emph{NOT} gate, a $|0\rangle$ qubit of
DNA\emph{pol} should also turn into $|1\rangle$ qubit, since
Pauli-$X$ transformation of controlled-\emph{NOT} gate corresponds
to a proton tunneling between the nucleobases and DNA\emph{pol}.
Therefore, value of the $\lambda_{K=3+k}$ remains fixed and
transformation $\mathbf{U}$ does not take any state $|s\rangle$ out
of the $DFS_{3+k}(2q-k+1)$ or $DFS_{3+k}(2q-k-1)$. This means that
decoherence is avoided during the formation of intrabase
entanglements.

Each state that corresponds to an intrabase entanglement (Equations
\ref{E-Q}, \ref{E-1}) lives in one of the DF subspaces $DFS_3(+1) =
\text{Span}\{|001\rangle,|010\rangle,|100\rangle\}$ and $DFS_3(-1) =
\text{Span}\{|011\rangle,|101\rangle,|110\rangle\}$. However,
intrabase entanglements should immune not only to the weak
collective decoherence, but also to the strong collective
decoherence: formation of the intrabase entanglement separates the
recognized nucleobase from the DNA\emph{pol}. This separation
removes the isolation provided by the enzyme and exposes the
nucleobase to the cellular environment. Assumption of the collective
decoherence to be weak may lose its validity by the removal of the
isolation. Thus, effects of the strong collective decoherence on the
intrabase entanglements should also be investigated. These effects
can be understood in terms of the actions of Pauli-$X$, -$Y$, and
-$Z$ transformations \citep{KE-36, KE-19}. $64 \times 64$ Pauli spin
matrices transform intrabase entanglements as follows:
\begin{eqnarray}
S_x: |\text{N}\rangle_{WC,Q} |\text{\={N}}\rangle_{WC,Q}
&\rightarrow& |\text{\={N}}\rangle_{WC,Q}|\text{N}\rangle_{WC,Q}
\\
S_y: |\text{N}\rangle_{WC,Q} |\text{\={N}}\rangle_{WC,Q}
&\rightarrow& |\text{\={N}}\rangle_{WC,Q}|\text{N}\rangle_{WC,Q}
\nonumber
\\
S_z: |\text{N}\rangle_{WC,Q} |\text{\={N}}\rangle_{WC,Q}
&\rightarrow& |\text{N}\rangle_{WC,Q}|\text{\={N}}\rangle_{WC,Q}
\nonumber
\end{eqnarray}

Since transformation $\mathbf{S}$ produces same interbase
entanglements from $|\text{N}\rangle_{WC,Q}
|\text{\={N}}\rangle_{WC,Q}$ and
$|\text{\={N}}\rangle_{WC,Q}|\text{N}\rangle_{WC,Q}$ states, effects
of the strong collective decoherence on the intrabase entanglements
seem to be trivial. This allows DNA\emph{pol} to safely search for
the complementary free nucleobase after the recognition of
nucleobase of ssDNA and to safely continue pairing of bases after
finishing the search.

Conservation of the proton number of nucleobase - DNA\emph{pol}
complex is trivial under the actions of swap gates and modified Bell
measurements. Hence, decoherence suppression during and after the
swapping protocol $\mathbf{S}$ can be demonstrated in a similar way
as is done for the transformation $\mathbf{U}$.

\section{Discussions}

In the presence of strong collective decoherence, smallest DFS (DF
subspace or subsystem) in which at least one qubit of information
can be encoded is a three qubit subsystem $DFS_{K=3}(J=1/2)$
\citep{KE-36, KE-19}. Construction of this subsystem involves the
use of four three-particle $J=1/2$ states. So, physical
implementation of the computation inside the smallest DFS requires
the use of four distinct building blocks each of which participates
in the computation over three physical qubits. Number of orthogonal
states living inside the smallest DFS imposes a restriction on the
number of different building block types, whereas required qubit
number for the computation inside the smallest DFS puts a limit on
the atom number present in the interaction region of these building
blocks.

Two of the four three-particle $J=1/2$ states that are constructed
by adding a two-particle $J=0$ state to a one-particle $J=1/2$ state
imply the entanglements of first two physical qubits of the
corresponding building blocks. Likewise, other two three-particle
$J=1/2$ states that are constructed by adding a two-particle $J=1$
state to a one-particle $J=1/2$ state imply the entanglements of all
the three physical qubits of the remaining building blocks. Also,
two states constructed by the same way should be associated in an
appropriate way to encode a DF qubit. The conventional way of DF
encoding inside the subsystem $DFS_{K=3}(J=1/2)$ is taking the
superpositions of these states. However, quantum superposition of
the corresponding building blocks are unlikely to be formed. Pairing
of the building blocks, whose entangled qubit numbers coincide,
would be a more reasonable way in a biological sense. Such a pairing
can be obtained by swapping intramolecular entanglements with
intermolecular entanglements. Since first two of the four $J=1/2$
states exclude the last physical qubits from the intramolecular
entanglements, intermolecular entanglements obtained by the pairing
of the corresponding building blocks should not include the
entanglement of their last physical qubits. On the contrary, pairing
of the other two building blocks should result in three
intermolecular entanglements between them.

In fact, if single-particle $|j=1/2, \, m_j=1/2\rangle$ state is
represented by $|0\rangle$ qubit, four three-particle $J=1/2$ states
are nothing else than the $|N\rangle_{WC,Q}$ states corresponding to
intrabase entanglements. It would be interesting if we could bring
the similarities between the discussed scenario and processing of
genetic information beyond an analogy.

Although tens of nucleotide derivatives are found in nature,
especially in tRNAs, genetic information is encoded by only four
nucleotides in the dsDNA. These deoxyribonucleotides usually form
Watson-Crick base pairs in which pairing occurs over the three $WC$
edge atoms of the nucleobases (Figure \ref{F-parts}): A and T form a
pair through two interbase hydrogen bonds, whereas G and C form a
pair through three interbase hydrogen bonds. The third atom in $WC$
edge of adenine is a carbon atom which is responsible for the lack
of one hydrogen bond in the A$\cdot$T pair. Preference of the carbon
rather than a more electronegative atom as the third $WC$ edge atom
of adenine may be a coincidental event stand out in the evolution of
nucleobases. Such a restriction on both nucleobase types and
interbase interactions used in the usual structure of the genetic
information should have an explanation in the evolutionary basis.

One of the essential problems of any information processing in
living systems is the presence of unavoidable noise. To survive,
organisms must have some special computational strategies for coping
with this problem. Since the nature of molecular realm is quantum
mechanical, making computation inside noiseless DFS may be a
favorable strategy. However, organisms must process information not
only more accurately, but also more powerfully. So, any discovery of
biological DF computation should have been followed by further
optimizations on the use of resources like energy, time, and number
of physical building blocks. Such a resource optimization actually
corresponds to a minimization of the qubit number required in the
computation. In this sense, restriction on both nucleobase types and
interbase interactions used in the usual structure of the genetic
code may have been a result of the further resource optimization in
biological DF computation.

\section{Conclusions}
\begin{figure*}[ht]
\centering
\includegraphics[width=0.78\textwidth]{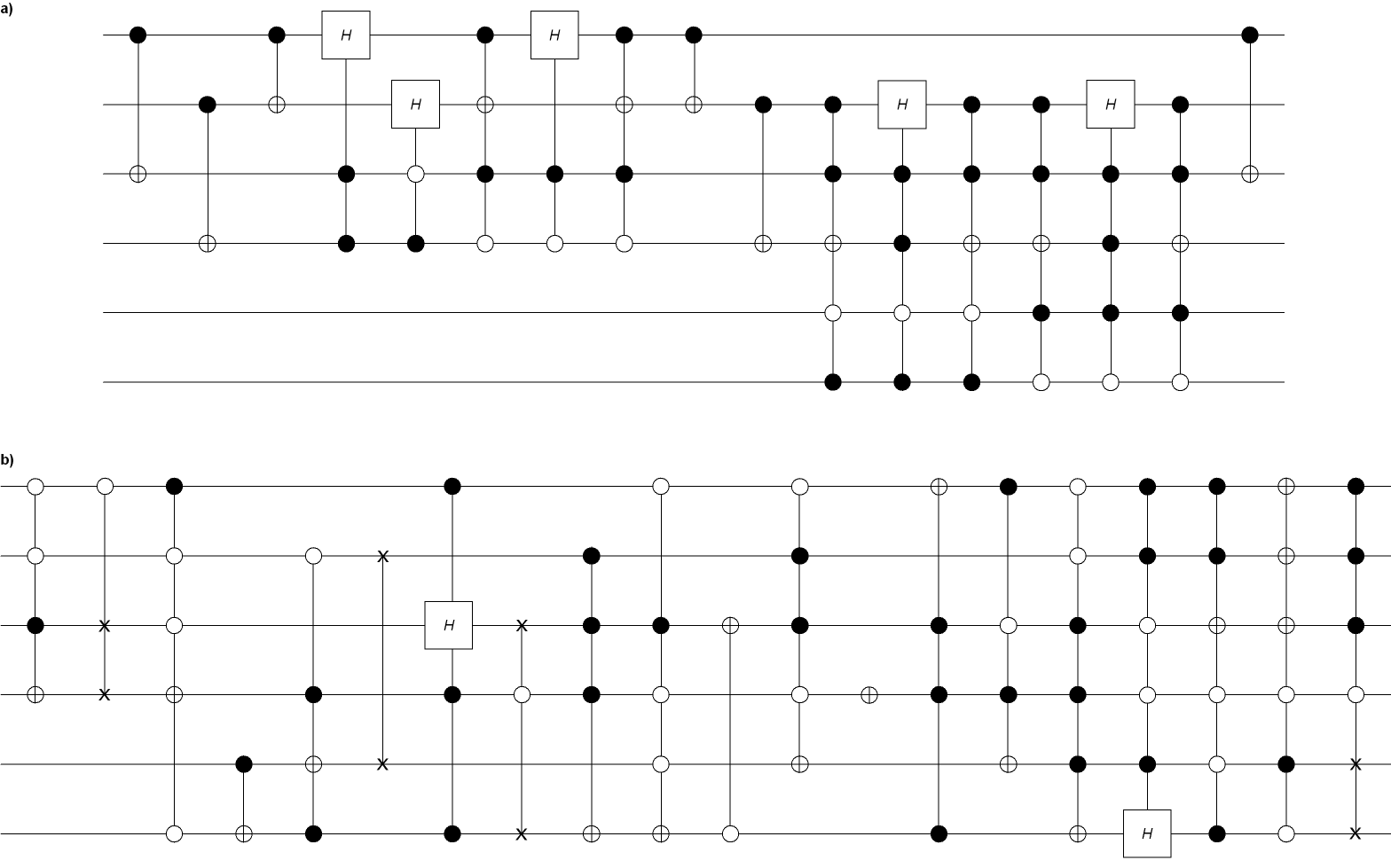}
\caption{Quantum transformation $V$ used in $\mathbf{S}$: quantum
gates indicated by a double $\times$ are swap gates. \emph{a}) First
part of the transformation $V$ which makes the states of the proper
base pairs orthogonal to each other and to the states of the
improper base pairs. \emph{b}) Second part of the transformation $V$
which converts the third qubit pair into $|01\rangle$ in G$\cdot$C
(or C$\cdot$G) pair and into $|11\rangle$ in A$\cdot$T (or
T$\cdot$A) pair. In the case of an improper base pairing, it
produces a superposition of states in which third qubit pair is
always $|00\rangle$ or $|10\rangle$.} \label{F-QC4}
\end{figure*}

Since all of the steps in both $\mathbf{U}$ and $\mathbf{S}$ can be
expressed as proton transfer and hydrogen bonding, the replication
scenario proposed in our model could be tested step by step with the
help of computational methods of quantum chemistry.

In addition to computational tests, some experimental setups may be
designed to explore some of the predictions of the model. For
example, states of base pairs immediately after the pairing are
obtained as in Equation \ref{E-4}. Indeed, $|\text{N}_1\cdot
\text{\={N}}_2\rangle_{WC,O}$ states are superpositions of different
tautomer forms with equal probability amplitudes. As discussed
before, these amplitudes should change by the quantum evolution in
the presence of the asymmetric double well potentials of the
hydrogen bonded atom pairs until they reach the actual values. Then,
it can be hypothesized that if this evolution can be prevented in a
proper way, probability of point mutations due to the formation of
rare tautomer forms should be higher than the ones obtained \emph{in
vivo}. This may be achieved by sufficiently decreasing the time
periods between two successive replications.

However, there are some deficiencies in the description of the
present model which should be removed before any computational or
experimental test. Most important deficiency is the absence of
enzyme's state in the computations. To obtain a more realistic
description, a state should be assigned to the enzyme and the whole
process including $\mathbf{U}$ and $\mathbf{S}$ should leave this
state invariant at the end. In \citeyear{ctc_1}, a self-consistency
condition for a quantum state was introduced to describe and
understand a disparate interaction \citep{ctc_1}. A similar
utilization of the self-consistent states in the description of the
enzymes seems to be appropriate. This is possible, even though the
state of an enzyme is expected to be mixed, and that the
purification of a mixed state in \citeauthor{ctc_1}'s formalism
\citep{ctc_1} is impossible \citep{ctc_6}. We note that proofs given
in \citep{ctc_6} are not valid for the mixed states which are used
to describe enzymes under the self-consistency condition. We plan to
investigate this further in the near future.

Finally, entanglement swapping may be a basic tool used by enzymes
and proteins in the cellular environment. If so, similar models may
be developed for amino acid - tRNA, aminoacyl-tRNA - mRNA, and amino
acid - amino acid interactions in the protein synthesis. If
successful models for these interactions can be developed, then we
can achieve a deeper understanding of the role of the quantum
effects and dynamics on the cellular information processing.
Perhaps, entanglement swapping will join and contribute to the
debates on the universal triplet genetic code \citep{KE-2, KE-14,
KE-15} and on the mechanism behind adaptive mutation \citep{KE-16,
KE-17, KE-18} after such models.

All in all, quantum effects are used mainly for the determination of
molecular shapes, sizes and chemical affinities in molecular biology
and biochemistry. Although functions of bio-molecules are explained
by structure, such as the complementary geometries of molecules and
weak intermolecular hydrogen bonds in nucleobase pairs, further
quantum effects are not thought to play any significant role in the
present biochemical complexity. However, they may be more useful
tools to understand the physics of life if quantum
circuits/protocols and organic molecules are considered as software
and hardware of the living systems. Reconsideration of evolution as
co-optimization of hardware (structure) and software (function),
reconciles two opposite approaches: natural selection and
self-organization. Thus, emergence of the life as a biochemical
complexity may be demystified in the context of quantum information
theory.

\section*{Acknowledgements}

Onur Pusuluk thanks Institute of Theoretical and Applied Physics
(ITAP) for the hospitality in the early stages of this work, and
would like to acknowledge support from the Scientific and
Technological Research Institute of Turkey (T\"{U}B\.{I}TAK)
National Scholarship Program for PhD Students.

\section*{Appendix A. Transformation $V$ used in $\mathbf{S}$}

Transformation $V$ used in $\mathbf{S}$ has thirty seven gates as
shown in Figure \ref{F-QC4}. Such a number seems to be very large
for an efficient replication process. However, Figure \ref{F-QC4} is
a general representation and all of the gates are not effectively
used in each base pair (see Table \ref{T-EBN}). In fact, average
effective gate number is approximately seventeen.
\begin{table}[h] \centering \caption{Effective gate
number of transformation $V$ (Figure \ref{F-QC4}) for each base pair
N$_1 \cdot$ \={N}$_2$.}
\begin{tabular}[c]{| c | p{0.1 cm} p{1 cm} p{1 cm} p{1 cm} p{0.7 cm} |}
\multicolumn{6}{c}{~} \\
\hline \backslashbox{N$_1$}{\\ \={N}$_2$} & & A & T & G & C
\\
\hline
A & & 12 & 11 & 20 & 19 \\
T & & 11 & 11 & 21 & 16 \\
G & & 17 & 16 & 22 & 22 \\
C & & 18 & 14 & 22 & 19 \\ \hline
\end{tabular}
\label{T-EBN}
\end{table}

\end{document}